\documentclass[aps,superscriptaddress,prc,twocolumn,nofootinbib]{revtex4}

\usepackage{graphicx}
\usepackage{epstopdf}
\usepackage{subfigure}
\usepackage{epsfig}
\usepackage{amsmath,amssymb,amsfonts}
\usepackage{color}

\begin{document}

\title{Modeling of heavy-flavor pair correlations in Au-Au collisions at 200$A$ GeV at the BNL Relativistic Heavy Ion Collider}

\author{Shanshan Cao}
\affiliation{Nuclear Science Division, Lawrence Berkeley National Laboratory, Berkeley, CA 94720, USA}
\affiliation{Department of Physics, Duke University, Durham, NC 27708, USA}

\author{Guang-You Qin}
\affiliation{Institute of Particle Physics and Key Laboratory of Quark and Lepton Physics (MOE), Central China Normal University, Wuhan, 430079, China}

\author{Steffen A. Bass}
\affiliation{Department of Physics, Duke University, Durham, NC 27708, USA}

\date{\today}


\begin{abstract}

We study the nuclear modification of angular and momentum correlations between heavy quark pairs in ultra-relativistic heavy-ion collisions. The evolution of heavy quarks inside the thermalized medium is described via a modified Langevin approach that incorporates both elastic and inelastic interactions with the medium constituents. The spacetime evolution of the fireball is obtained from a (2+1)-dimensional viscous hydrodynamics simulation. The hadronization of heavy quarks is performed utilizing a hybrid model of fragmentation and coalescence. Our results show that the nuclear modification of the transverse momentum imbalance of $D\overline{D}$ pairs reflects the total energy loss experienced by the heavy quarks and may help us probe specific regions of the medium. The angular correlation of heavy flavor pairs, especially in the low to intermediate transverse momentum regime, is sensitive to the detailed energy loss mechanism of heavy quarks inside the QGP.

\end{abstract}

\maketitle


\section{Introduction}
\label{sec:introduction}

It has been shown that a new state of matter, commonly referred to as the strongly interacting quark-gluon plasma (sQGP), can be produced in ultra-relativistic nucleus-nucleus collisions, such as those performed at the CERN Large Hadron Collider (LHC) and the BNL Relativistic Heavy-ion Collider (RHIC). This highly-excited nuclear matter exhibits many remarkable properties \cite{Gyulassy:2004zy,Adcox:2004mh,Arsene:2004fa,Back:2004je,Adams:2005dq,Muller:2006ee,Muller:2012zq,Qin:2015nma}, among most important observations is the strong anisotropic collective flow exhibited by the hot and dense fireball and the final emitted hadrons \cite{Adare:2011tg, Adamczyk:2013waa, ATLAS:2012at}, which can be nicely described by relativistic hydrodynamics simulations \cite{Heinz:2013th, Gale:2013da, Song:2013gia, Romatschke:2009im, Huovinen:2013wma}. Another result is the strong modification of large transverse momentum ($p_\mathrm{T}$) hadrons and jets as compared to the reference of proton-proton collisions \cite{Adcox:2001jp,Adler:2002xw,Aamodt:2010jd}. Many high-$p_\mathrm{T}$ phenomena and jet-related observables can be well understood from parton energy loss which originates from the interaction between the propagating hard partons and the hot and dense QCD matter \cite{Bass:2008rv, Burke:2013yra}.

Heavy quarks and heavy flavor mesons are of particular importance.
Because of their large masses, heavy quarks are expected to lose less energy than light flavor partons due to the so-called ``dead cone" effect, i.e., the phase space of collinear radiation is suppressed by the kinematics compared to light partons \cite{Dokshitzer:2001zm}.
In contrast, experimental measurements have shown that the nuclear modification factors $R_\mathrm{AA}$ and the anisotropy coefficients $v_2$ for heavy flavor mesons are comparable to light flavor hadrons \cite{Adare:2010de,Adare:2014rly,Tlusty:2012ix,Adamczyk:2014uip,Grelli:2012yv,Abelev:2013lca}.

Many phenomenological models have been developed to investigate the propagation and energy loss of heavy quarks inside the hot and dense nuclear medium \cite{Moore:2004tg,Gossiaux:2010yx,He:2011qa,Young:2011ug,Uphoff:2012gb,Nahrgang:2013saa,Cao:2013ita,Cao:2015hia}.
In our earlier work \cite{Cao:2011et, Cao:2012jt, Cao:2012au, Cao:2013ita,Cao:2015hia}, we have developed a transport model to simulate the evolution dynamics of heavy flavors in relativistic heavy-ion collisions. 
Both collisional and radiative energy losses of heavy quarks are included in our model via an improved Langevin approach, where the gluon radiation is treated as a recoil force exerted on the propagating heavy quarks.
The recoil force is obtained from the medium-induced gluon radiation spectrum as calculated from perturbative QCD energy loss formalisms. 
The hadronization process of heavy quarks after passing through the QGP is simulated utilizing a hybrid model of fragmentation and coalescence, where the momentum dependence of the relative probability of the two hadronization mechanisms is calculated according to the Wigner functions of the heavy hadrons. The interactions of heavy flavor mesons with the hadron gas were also incorporated in Ref. \cite{Cao:2015hia} by utilizing the Ultra-relativistic Quantum Molecular Dynamic Model (UrQMD) \cite{Bass:1998ca}. 
With the above setup, we can provide a reasonable description of the nuclear modification of heavy flavor mesons at the LHC and RHIC \cite{Cao:2015hia}.

In addition to single inclusive observables, correlations of final state particles have played a very important role in the study of jet quenching and parton energy loss. 
In the light flavor sector, there have been a wealth of experimental measurements and theoretical studies on final state correlations, such as back-to-back dihadron and photon-hadron correlations \cite{Adler:2002tq, Adams:2006yt, Adare:2009vd, Abelev:2009gu,Zhang:2007ja,Qin:2009bk}, as well as the momentum imbalance of back-to-back jet pairs and photon-jet pairs \cite{Aad:2010bu, Chatrchyan:2011sx, Aad:2014bxa, Qin:2010mn, Dai:2012am, Qin:2012gp, Wang:2013cia}.
In the heavy flavor sector, the angular correlations of heavy flavor pairs have been studied in Ref. \cite{Zhu:2006er,Zhu:2007ne,Gossiaux:2006yu,Akamatsu:2009ya,Younus:2013be,Nahrgang:2013saa,Cao:2013wka,Cao:2014pka,Cao:2014dja}, and the momentum imbalance of heavy flavor pairs has been investigated in Ref. \cite{Gossiaux:2009mk,Uphoff:2013rka}. 
In this article, we present a detailed study of both angular and momentum correlations for heavy flavor pairs using our heavy flavor transport model built in Ref. \cite{Cao:2011et, Cao:2012jt, Cao:2012au, Cao:2013ita}. 
The dependence on the collisions centrality and the transverse momentum cuts are investigated for both angular and momentum correlations.
We will show that while the momentum imbalance of back-to-back heavy flavor meson pairs is mostly sensitive to the total energy loss of heavy quarks in the dense QGP medium, the angular correlations probe the transverse momentum broadening of heavy quark propagation and are sensitive to different parton energy loss mechanisms, especially in the low transverse momentum region.

This article is organized as follows. In Sec.\ref{sec:HQDynamics}, we present our theoretical framework for simulating the in-medium evolution and hadronization of heavy quarks in relativistic heavy-ion collisions. In Sec.\ref{sec:correlation}, we provide a detailed investigation on the angular correlations of heavy flavor pairs and show their sensitivity to different energy loss mechanisms. The transverse momentum imbalance between heavy flavor pairs is presented in Sec.\ref{sec:pAsym}. Sec.\ref{sec:conclusions} contains our conclusions.

\section{Heavy flavor dynamics}
\label{sec:HQDynamics}

In the limit of multiple scatterings, the momentum exchange of heavy quarks in each scattering with the medium constituents is small. In this case, the evolution of heavy quarks inside a thermalized medium via multiple quasi-elastic scatterings can be described by the Langevin equation. However, as one moves to the high $p_\mathrm{T}$ region, quasi-elastic scatterings alone are no longer sufficient to describe the heavy quark evolution and energy loss in hot and dense nuclear medium, and the contribution from medium-induced gluon radiation has to be included. 
To incorporate both energy loss processes, we follow our previous framework \cite{Cao:2013ita,Cao:2015hia} and modify the classical Langevin equation as follows:
\begin{equation}
\label{eq:modifiedLangevin}
\frac{d\vec{p}}{dt}=-\eta_D(p)\vec{p}+\vec{\xi}+\vec{f}_g.
\end{equation}
In the above equation, the first two terms on the right hand side are the drag force and the thermal random force experienced by heavy quarks when they scatter with the medium constituents; they are inherited from the classical Langevin equation.  In this work, we assume the fluctuation-dissipation theorem still holds, $\eta_D(p)=\kappa/(2TE)$, where $\kappa$ is the momentum space diffusion coefficient of heavy quark defined as $\langle\xi^i(t)\xi^j(t')\rangle=\kappa\delta^{ij}\delta(t-t')$. On the right hand of the above equation, a third term $\vec{f}_g$ is introduced to describe the recoil force experienced by heavy quarks when they radiate gluons. 
The probability for a heavy quark to emit a gluon within the time interval $[t,t+\Delta t]$ is evaluated according to the average number of radiated gluons:
\begin{equation}
\label{eq:gluonnumber}
P_\mathrm{rad}(t,\Delta t)=\langle N_\mathrm{g}(t,\Delta t)\rangle = \Delta t \int dxdk_\perp^2 \frac{dN_\mathrm{g}}{dx dk_\perp^2 dt},
\end{equation}
where $\Delta t$ is chosen to be sufficiently small to ensure $P_\mathrm{rad}(t,\Delta t)<1$. 
${dN_\mathrm{g}}/{dx dk_\perp^2 dt}$ is the gluon radiation spectrum and in our work is obtained from the higher-twist energy loss calculation \cite{Guo:2000nz,Majumder:2009ge,Zhang:2003wk}:
\begin{equation}
\label{eq:gluondistribution}
\frac{dN_\mathrm{g}}{dx dk_\perp^2 dt}=\frac{2 \alpha_s  P(x)\hat{q} }{\pi k_\perp^4} {\sin}^2\left(\frac{t-t_i}{2\tau_f}\right)\left(\frac{k_\perp^2}{k_\perp^2+x^2 M^2}\right)^4.
\end{equation}
In the equation, $x$ is the fractional energy carried by the radiated gluon, $k_\perp$ is its transverse momentum, $\tau_f={2Ex(1-x)}/{(k_\perp^2+x^2M^2)}$ is its formation time and $P(x)$ is the vacuum splitting function. $\hat{q}$ is known as the gluon transport coefficient, which is related to $\kappa$ via $\hat{q}=2 \kappa C_A / C_F$. With this setup, there is only one free parameter in our model. In the literature, one often quotes the spatial diffusion coefficient $D$ of heavy quark to characterize its interaction strength with the medium; it is related to the momentum drag and diffusion as: $D\equiv T/[M\eta_D(0)]=2T^2/\kappa$. We will follow such practice and quote the values of $D$ in the following discussions.

Using our transport model, we can simulate the propagation of heavy quarks inside the QGP matter created in relativistic heavy-ion collisions. The spacetime evolution of the local temperature and fluid velocity of the dense nuclear matter is simulated with a (2+1)-dimensional viscous hydrodynamic model \cite{Qiu:2011hf}. In each time step, we boost the heavy quark into the local rest frame of the fluid cell in which the energy and momentum of a given heavy quark is updated according to our modified Langevin equation before it is boosted back to the global (laboratory) frame.

Before the start of the hydrodynamical evolution (at $\tau_0=0.6$~fm/c), the initial entropy density is obtained via a Monte-Carlo (MC) Glauber model. 
The same MC Glauber model is utilized to initialize the spatial distribution of initial produced heavy quarks. 
The initial momentum space distribution of heavy quarks is calculated in leading order perturbative QCD when calculating the production and nuclear modification of single inclusive heavy flavor mesons. The nuclear shadowing effect in nucleus-nucleus collisions has been included by utilizing the EPS09 parametrization \cite{Eskola:2009uj}. In the study of heavy flavor correlations, the initial angular and transverse momentum distributions of heavy flavor pairs are obtained from \textsc{Pythia} 8 simulation. 
The evolution of heavy quarks inside the QGP ends when they reach fluid cells with local temperature below $T_\mathrm{c}$ (165~MeV). On the freeze-out hypersurface, heavy quarks hadronize into color neutral bound states using our hybrid model of fragmentation and coalescence developed in Ref. \cite{Cao:2013ita,Cao:2015hia}. 
In such a hybrid model, high momentum heavy quarks tend to fragment directly into hadrons, but low momentum heavy quarks are more probable to combine with thermal partons from the medium to form heavy flavor mesons. The relative probability between these two processes is determined by the Wigner functions based on an instantaneous coalescence model \cite{Oh:2009zj}. Within our coalescence model, the momentum distribution of the produced hadron can be calculated once a heavy quark is chosen to recombine with a thermal parton. If the heavy quark is selected to fragment, \textsc{Pythia} 6 \cite{Sjostrand:2006za} is used to simulate the fragmentation process using Peterson fragmentation function. The rescattering of the produced heavy mesons in a hadron gas will not be included in this work, since our previous study \cite{Cao:2015hia} has shown that its effect can be reasonably approximated by slightly increasing the momentum space transport coefficient of heavy quarks in the QGP phase.

\begin{figure}[tb]
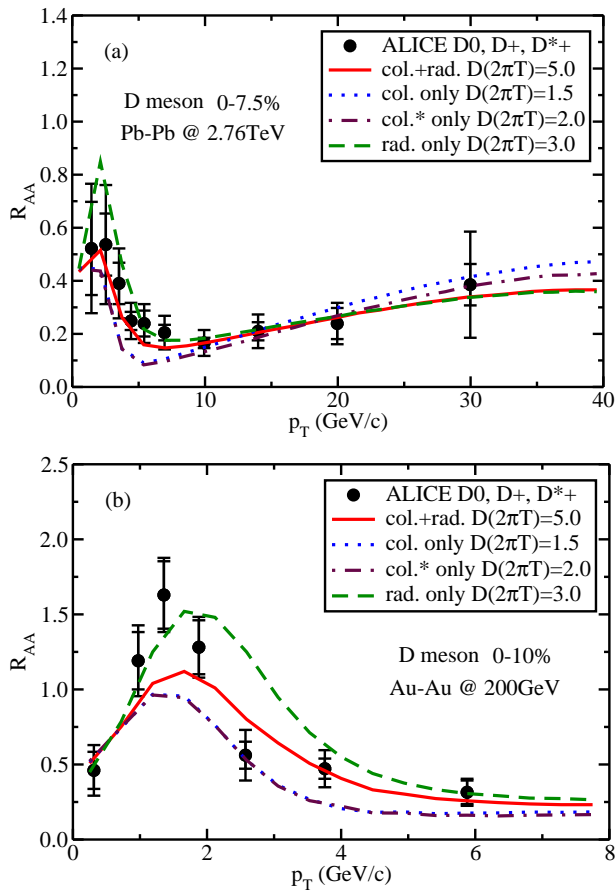

  \subfigure{\label{fig:plot-fitLHCRAAforCorr}
      \epsfig{file=plot-fitLHCRAAforCorr.eps, width=0.45\textwidth, clip=}}
  \subfigure{\label{fig:plot-fitRHICRAAforCorr}
      \epsfig{file=plot-fitRHICRAAforCorr.eps, width=0.45\textwidth, clip=}}
  \caption{(Color online) $D$ meson nuclear modification factor $R_{AA}$ in (a) central Pb-Pb collisions and (b) central Au-Au collisions. Different energy loss mechanisms are used for comparison.}
 \label{fig:plot-fitRAAforCorr}
\end{figure}

In our model, the transport coefficient of heavy quarks is determined by comparing to the experimental data on heavy meson nuclear modification at high $p_\mathrm{T}$ and then applied to calculations at lower $p_\mathrm{T}$ and other heavy flavor observables. In Fig. \ref{fig:plot-fitRAAforCorr}, we provide calculations of the $D$ meson $R_\mathrm{AA}$ in 2.76 TeV central Pb-Pb collisions and 200 GeV Au-Au collisions. Different energy loss mechanisms are utilized and compared to the experimental data. We can see that with the simultaneous inclusion of  collisional and radiative processes, the use of $D=5/(2\pi T)$ provides the best description for the $p_\mathrm{T}$ dependence of the $D$ meson $R_\mathrm{AA}$. If only quasi-elastic scattering or only gluon radiation are included, one can also obtain a reasonable description of the experimental data if the diffusion coefficient is adjusted to $D=1.5/(2\pi T)$ or $D=3/(2\pi T)$, respectively. The uncertainty introduced by a different velocity dependence of the longitudinal and transverse transport coefficients ($\kappa_\mathrm{L}$ and $\kappa_\mathrm{T}$) for the collisional energy loss (denoted by the star notation) is investigated as well: the velocity dependence calculated to the leading logarithm is taken from Ref. \cite{Moore:2004tg}. In Fig. \ref{fig:plot-fitRAAforCorr} we show that an adjustment of the diffusion constant to $D=2/(2\pi T)$ is able to reproduce our earlier results obtained with velocity-independent transport coefficients. Fig. \ref{fig:plot-fitRAAforCorr} implies that the single particle spectrum appears not sufficient to distinguish between the contributions from different energy loss mechanisms. In the next section, we will show that this large uncertainty can be reduced by investigating the correlations of heavy flavor pairs. In this work we perform the calculations for 200~GeV Au-Au collisions at RHIC.

\section{Angular correlations of heavy flavor pairs}
\label{sec:correlation}

In this section, we study the angular correlations of heavy flavor pairs and their nuclear modifications in relativistic heavy-ion collisions. The dependence of the nuclear modification effect on the centrality cut and the transverse momentum $p_\mathrm{T}$ cuts will be investigated in some details. 


\begin{figure}[tb]
  \epsfig{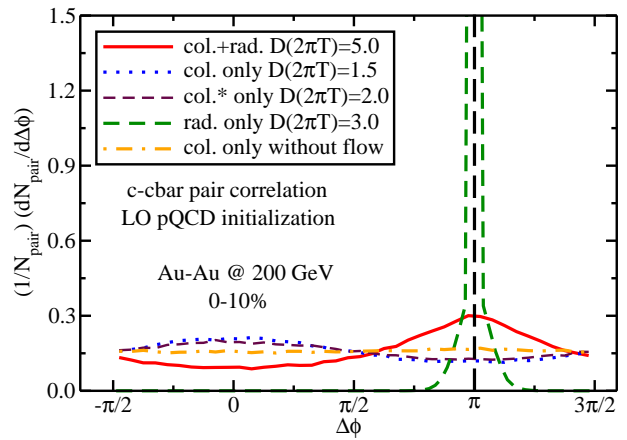}
  \caption{(Color online) Angular correlations of $c\overline{c}$ pairs in central Au-Au collisions. Here the LO pQCD back-to-back approximation is used for $c\overline{c}$ pair initialization.}
 \label{fig:plot-corr-ccbar-00-10-00pt}
\end{figure}

As a first academic study, we initialize heavy quark pair production using LO pQCD calculation, where $c$ and $\overline{c}$ pairs are assumed to be back-to-back and have the same magnitudes of transvese momentum. 
We calculate the angular correlation functions of $c\overline{c}$ pairs after they traverse a realistic QGP medium created in central Au-Au collisions at RHIC. The result is shown in Fig. \ref{fig:plot-corr-ccbar-00-10-00pt}, in which $N_\mathrm{pair}$ denotes the number of $c\overline{c}$ pairs (later for $D-\overline{D}$ pairs within the selected trigger class) so that our correlation function is normalized to unity. The same values of transport coefficients are used here as in Fig. \ref{fig:plot-fitRAAforCorr}. 
We can see that while different energy loss mechanisms provide similar amount of suppression of single $D$ mesons, they result in very different angular correlation functions of $c\overline{c}$ pairs in the final state. 
Since the medium-induced gluon radiation is mostly dominated by the transverse momentum kicks exerted on the radiated gluons, the effect on the propagation directions of the heavy quarks is very mild. 
One can see that the angular distribution of the final state $c\overline{c}$ pairs still peaks around $\Delta\phi = \pi$ if only radiative energy loss is included. 
However, quasi-elastic scatterings are very effective in changing the heavy quark propagation directions, and the final angular distribution is widely spread out. 
Such result implies that the angular correlations of heavy flavor pairs provide a possibility to distinguish different energy loss mechanisms.

One interesting observation in Fig. \ref{fig:plot-corr-ccbar-00-10-00pt} is that purely collisional energy loss leads to a peak around $\Delta\phi = 0$ in the angular distribution. This indicates that a lot of low momentum $c\overline{c}$ pairs move collinearly in the final state even though they are initialized as back-to-back pairs. 
This is caused by the strong radial flow of the hydrodynamic background which boosts the $c\overline{c}$ pairs into its own expanding direction. Such effect was historically referred to as the ``partonic wind effect" in Ref. \cite{Zhu:2007ne}; it might lead to some enhancement of quarkonium regeneration in a QGP matter. We verify this effect by solving the Langevin equation in the global (laboratory) frame, i.e., the medium flow effect is shut off manually and only the temperature profile affects the heavy quark evolution. With the flow effect turned off, the peak around $\Delta\phi = 0$ disappears and the angular distribution of $c\overline{c}$ pairs becomes flat.
The flat distribution also means that the final charm quark propagation directions can be entirely randomized by the strong quasielastic scatterings inside the QGP. In Fig. \ref{fig:plot-corr-ccbar-00-10-00pt}, we also observe that collisional energy loss with both velocity independent and velocity dependent $\kappa_\mathrm{L}$ and $\kappa_\mathrm{T}$ lead to similar angular correlation functions as long as the diffusion coefficients are properly adjusted to describe the observed $D$ meson $R_\mathrm{AA}$ as shown in Fig. \ref{fig:plot-fitRAAforCorr}. In the rest of this work, we will adopt the velocity independent $\kappa$ as assumed in our earlier study \cite{Cao:2015hia}.


\begin{figure}[tb]
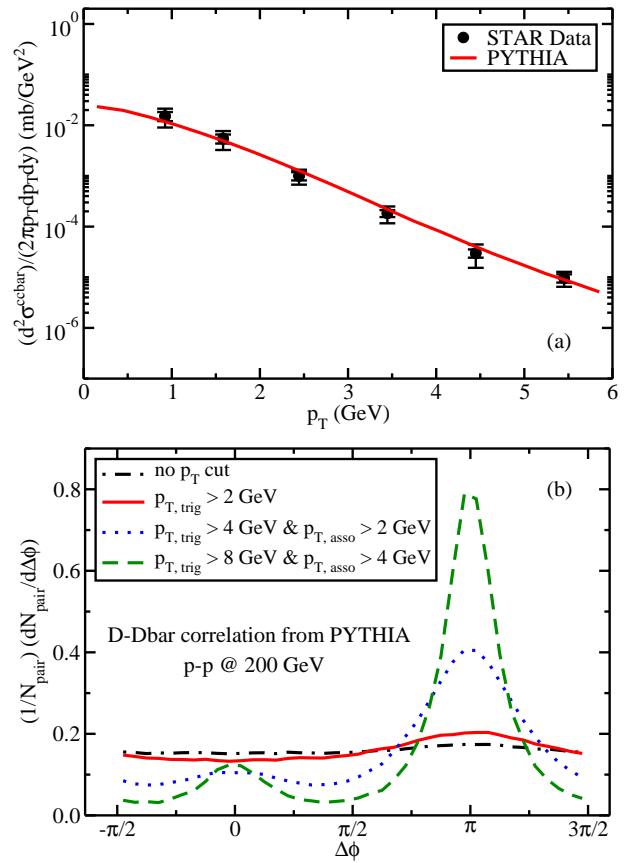

 \subfigure{\label{fig:plot-single-c-initial}
      \epsfig{file=plot-single-c-initial.eps, width=0.45\textwidth, clip=}}
 \subfigure{\label{fig:plot-corr-DDinit-angle}
      \epsfig{file=plot-corr-DDinit-angle.eps, width=0.44\textwidth, clip=}}
 \caption{(Color online) (a) $p_\mathrm{T}$ distribution of single inclusive $D$ mesons, and (b) the angular correlations of $D\overline{D}$ pair in proton-proton collisions at RHIC obtained from \textsc{Pythia} 8 simulation.}
 \label{fig:pythiaInitial}
\end{figure}

The above study of the correlations between $c\overline{c}$ pairs is only for academic purposes. For the final state correlations of $D\overline{D}$ pairs, we first note that there might be more than one $c\overline{c}$ pair produced in one single event of heavy-ion collisions (which depends on the kinematic cuts and collision centrality). To take this effect into account, for each event we pair every $D$ meson with all $\overline{D}$ mesons, and analyze all final state $D\overline{D}$ pairs. It is clear that there are uncorrelated $D\overline{D}$ pairs (background) contributing to the final correlations, especially with low $p_\mathrm{T}$ cuts and in more central collisions. 

For a more realistic simulation, \textsc{Pythia} 8 is utilized to simulate the initial $c\overline{c}$ pair production. We adopt the tunings of the \textsc{Pythia} 8 generator \cite{Shi:2015cva} that reproduce the spectrum of single charm hadrons in proton-proton collisions at RHIC as measured by the STAR Collaboration \cite{Adamczyk:2012af} [see Fig. \ref{fig:plot-single-c-initial}], and then use the same parameter setting to obtain the angular correlation functions of $D\overline{D}$ pairs in proton-proton collisions as shown in Fig. \ref{fig:plot-corr-DDinit-angle}. We can see that the $D\overline{D}$ angular correlations are pretty flat if no transverse momentum cut is applied. If one imposes a 2~GeV $p_\mathrm{T}$ cut for the trigger $D$ or $\overline{D}$ meson, a peak appears around $\Delta\phi = \pi$. With higher and higher $p_\mathrm{T}$ cuts applied for the trigger and associated heavy mesons, the away side peak becomes more and more prominent. Another interesting observation is that with a high enough $p_\mathrm{T}$ cut, a near-side peak begin to appear; this is the contribution from the initial gluon splitting process. The above obtained results serve as a baseline for the following study of $D\overline{D}$ correlations in Au-Au collisions at RHIC.

\begin{figure}[tb]
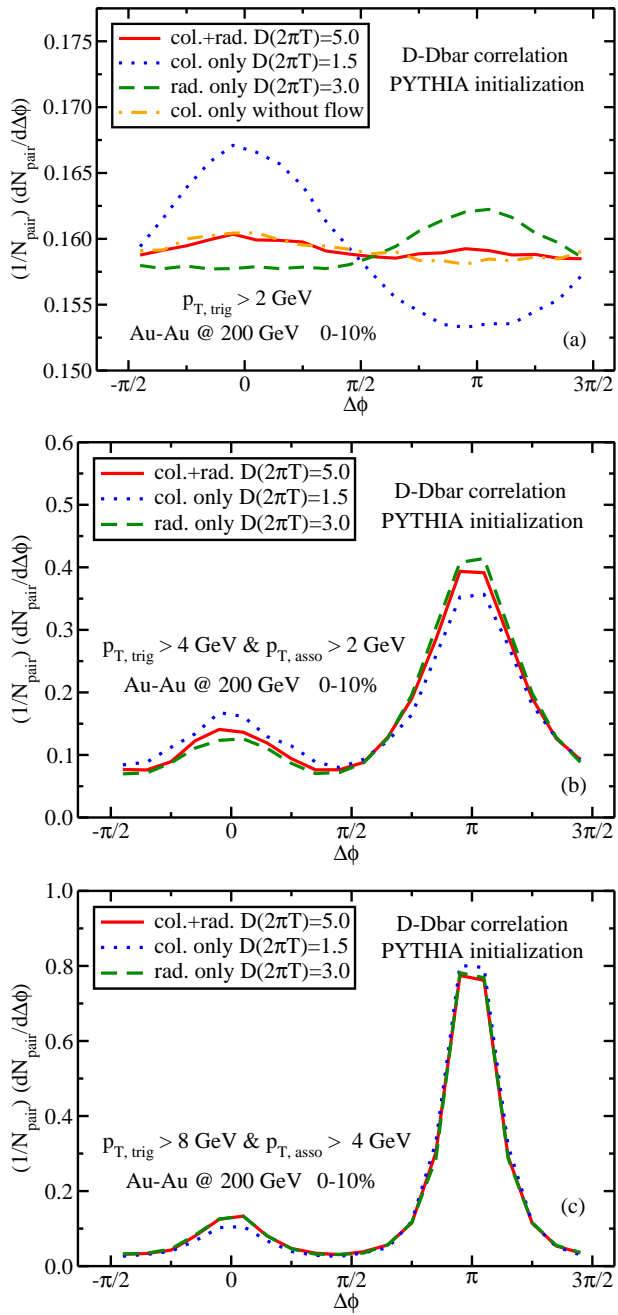

 \subfigure{\label{fig:plot-corr-DDbar-00-10-20pt}
      \epsfig{file=plot-corr-DDbar-00-10-20pt.eps, width=0.45\textwidth, clip=}}
 \subfigure{\label{fig:plot-corr-DDbar-00-10-42pt}
      \epsfig{file=plot-corr-DDbar-00-10-42pt.eps, width=0.45\textwidth, clip=}}
 \subfigure{\label{fig:plot-corr-DDbar-00-10-84pt}
      \epsfig{file=plot-corr-DDbar-00-10-84pt.eps, width=0.45\textwidth, clip=}}
 \caption{(Color online) $D$-$\overline{D}$ angular correlations in central Au-Au collisions at RHIC. Different energy loss mechanisms and different $p_\mathrm{T}$ cuts are shown for comparison.}
 \label{fig:ang-DD}
\end{figure}

In Fig. \ref{fig:ang-DD}, we show the angular correlation functions of $D\overline{D}$ pairs for 200~GeV central Au-Au collisions at RHIC. Fig. \ref{fig:plot-corr-DDbar-00-10-20pt} shows the result when a 2~GeV $p_\mathrm{T}$ cut is applied to the trigger $D/\overline{D}$ meson. Similar to $c\overline{c}$ pair correlation shown in Fig. \ref{fig:plot-corr-ccbar-00-10-00pt}, the purely radiative energy loss does not strongly affect the angular correlation very much [comparing the dashed line in Fig. \ref{fig:plot-corr-DDbar-00-10-20pt} and the solid line in Fig. \ref{fig:plot-corr-DDinit-angle}]. However, purely collisional energy loss is very effective in suppressing the away-side peak. Elastic scatterings also produce a near side peak around $\Delta\phi =0$, this effect diminishes when the coupling between the heavy quarks and the collective flow of the QGP is switched off. The realistic scenario in which a combination of both energy loss mechanisms is enabled falls between the above two curves. While these findings may help distinguish between different energy loss mechanisms, note that systematic uncertainties exist for the radiative energy loss at low $p_\mathrm{T}$ since currently the detailed balance between gluon emission and absorption is not implemented and we employ a low momentum cut-off below which heavy quarks can only interact via elastic scattering to retain detailed balance for heavy quarks that have thermalized with the bulk QCD medium. Another feature of $D\overline{D}$ correlations is the large background contributed from uncorrelated $D\overline{D}$ pairs, as compared to $c\overline{c}$ angular correlations. In our calculation, there are on average around 10 pairs of $c\overline{c}$'s produced in each central Au-Au collision; the background contribution diminishes when one moves to peripheral collisions or larger $p_\mathrm{T}$ cuts are applied. For example, $\langle N_\textrm{pair} \rangle$ drops below 1 per trigger event in Fig. \ref{fig:plot-corr-DDbar-00-10-20pt} when the centrality is greater than about 50\%.
Fig. \ref{fig:plot-corr-DDbar-00-10-42pt} shows the same results, but with the application of a 4~GeV $p_\mathrm{T}$ cut for the trigger $D$ or $\overline{D}$ meson and a 2~GeV cut for the associated meson. 
One can see that the difference between collisional and radiative energy losses becomes smaller than that in Fig. \ref{fig:plot-corr-DDbar-00-10-20pt}. The difference almost disappears with higher enough $p_\mathrm{T}$ cuts applied, see e.g., Fig. \ref{fig:plot-corr-DDbar-00-10-84pt}. As shown by the initial spectrum of charm quarks at RHIC energy [Fig. \ref{fig:plot-single-c-initial}], with the $p_\mathrm{T}$ cuts applied in Fig. \ref{fig:plot-corr-DDbar-00-10-42pt} and \ref{fig:plot-corr-DDbar-00-10-84pt}, on average there should be less than one pair of $c\overline{c}$ produced in each event, and thus very few uncorrelated $D\overline{D}$ pairs contribute to these two scenarios.

\begin{figure}[tb]
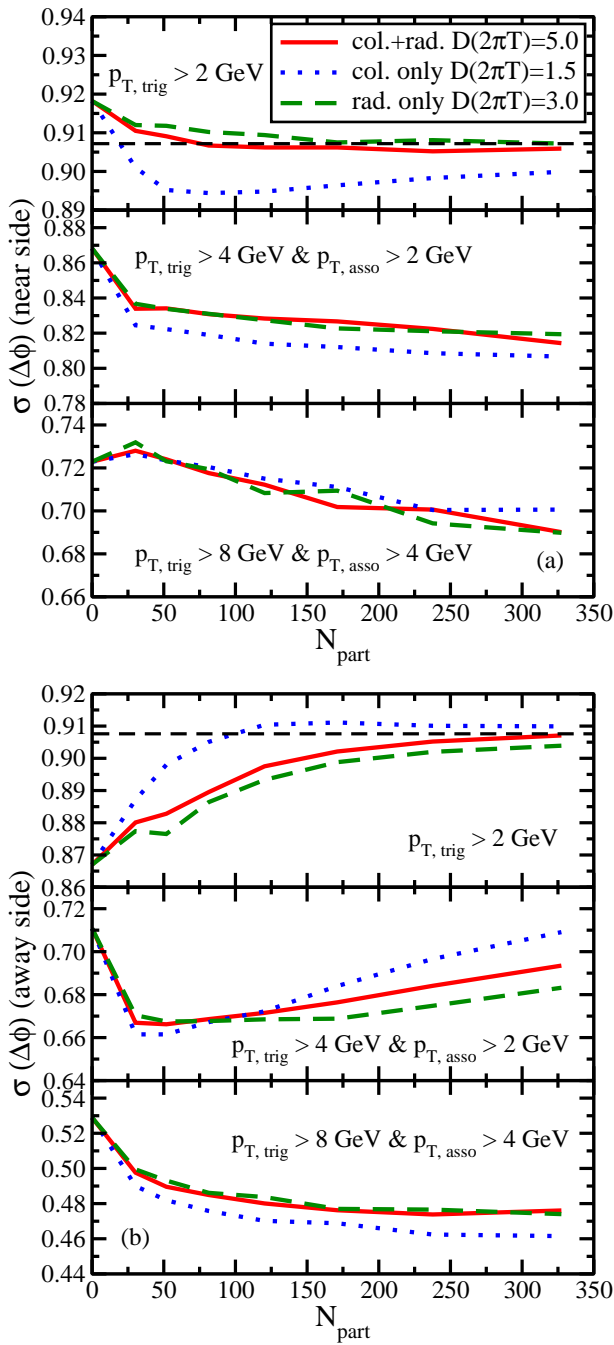

 \subfigure{\label{fig:plot-var1-DDbar-npart}
      \epsfig{file=plot-var1-DDbar-npart.eps, width=0.45\textwidth, clip=}}
 \subfigure{\label{fig:plot-var2-DDbar-npart}
      \epsfig{file=plot-var2-DDbar-npart.eps, width=0.45\textwidth, clip=}}
 \caption{(Color online) Variances of (a) near-side and (b) away-side distribution in $D\overline{D}$ angular correlation function as functions of the participant number. Different energy loss mechanisms and different $p_\mathrm{T}$ cuts are shown for comparison.}
 \label{fig:var-DD}
\end{figure}

To further quantify the medium modification of the angular correlations of heavy flavor meson pairs, we calculate the variances of the angular distributions for both near side $(-\pi/2,\pi/2)$ and away side $(\pi/2,3\pi/2)$.
The result is shown in Fig. \ref{fig:var-DD} where the variances are plotted as functions of the participant numbers $N_{\rm part}$ in Au-Au collisions at RHIC. The points at $N_\mathrm{part}=0$ correspond to the baseline proton-proton collisions. Note that the variance of $\Delta \phi$ is defined as $\left\langle(\Delta\phi-\langle\Delta\phi\rangle)^2\right\rangle=\langle\Delta\phi^2\rangle-\langle\Delta\phi\rangle^2$. And for the purpose of a better illustration, we have drawn a horizontal dashed line at $\pi/\sqrt{12}$ which is the variance for a uniform distribution in $(-\pi/2,\pi/2)$ or in $(\pi/2,3\pi/2)$; a smaller variance than $\pi/\sqrt{12}$ implies a peak around $\Delta\phi = 0$ or $\Delta\phi = \pi$, and a larger variance than $\pi/\sqrt{12}$ implies a dip on the near or away side.

Figure \ref{fig:plot-var1-DDbar-npart} shows the variances for the near side correlations. When a 2~GeV cut is applied to the trigger $p_\mathrm{T}$ (the upper panel), there is a dip structure in the $D\overline{D}$ correlations in proton-proton collisions and peripheral Au-Au collisions (therefore, the variance is larger than the value of a uniform distribution). With purely radiative energy loss, such dip gradually approaches a uniform distribution as one moves to more central collisions. However, with purely collisional energy loss, the variance first drops below $\pi/\sqrt{12}$ when $N_\mathrm{part}$ is not very large, and then increases and gradually approaches to the uniform distribution value as one moves to more central collisions. The smaller variance than $\pi/\sqrt{12}$ in central and mid-central collisions indicates there is a peak formed around $\Delta \phi=0$ due to the ``partonic wind effect" as discussed earlier. The non-monotonic behavior of the variance as a function of $N_{\rm part}$ originates from the increasing contribution from the uncorrelated $D\overline{D}$ pairs in more central collisions which smoothes out the angular distribution and make the variance approach the uniform distribution value.

With the application of larger $p_\mathrm{T}$ cuts [the middle and lower panels in Fig. \ref{fig:plot-var1-DDbar-npart}], the variance at the near side starts from a value below $\pi/\sqrt{12}$ for proton-proton collisions and decreases with the increase of $N_\mathrm{part}$ in Au-Au collisions. 
This behavior can be understood from the in-medium energy loss of heavy quarks as follows.
The final $D/\overline{D}$ mesons within certain $p_\mathrm{T}$ cuts originate from initial $c/\overline{c}$ quark which have much higher momentum and thus smaller variances for their angular distributions [see Fig. \ref{fig:plot-corr-DDinit-angle}]. 
Since the angular correlations do not change very much at high $p_\mathrm{T}$ after in-medium evolution [see e.g., Fig. \ref{fig:plot-corr-DDbar-00-10-84pt}], one obtains more peaked angular distribution for heavy flavor pairs for more central collisions. 
We further verify this by applying the transverse momentum cuts to the initial heavy quarks instead of the final observed heavy mesons. 
For such fixed sample of heavy quark/meson pairs, we do find that the variance of the angular distribution increases monotonically towards $\pi/\sqrt{12}$ when increasing $N_{\rm part}$. 
Again, we find that the difference among different energy loss mechanisms is larger for low $p_\mathrm{T}$ meson pairs and becomes smaller for larger $p_\mathrm{T}$ meson pairs.

The variances for the away-side angular distribution are shown in Fig. \ref{fig:plot-var2-DDbar-npart}. 
With small $p_\mathrm{T}$ cut (2~GeV), purely radiative energy loss smoothes the away-side peak in $D\overline{D}$ correlation distribution and the variance monotonically approaches towards the uniform distribution value. However, with purely collisional energy loss, the variance first increases above the uniform distribution limit which indicates a depletion of $D\overline{D}$ pairs around $\Delta\phi = \pi$, and then decreases towards $\pi/\sqrt{12}$ due to the smoothing of the background contribution.
With higher $p_\mathrm{T}$ cuts, there is the interplay between two combinational effects: the transverse momentum broadening and the energy loss of heavy quarks inside the QGP. 
The former increases the variance towards the uniform distribution limit and is more efficient for relatively lower $p_\mathrm{T}$, while the latter enhances the contribution from the charm quarks with higher initial $p_\mathrm{T}$ which reduces the variance of the final angular distribution. 
As a consequence, a non-monotonic behavior of the variance is observed as a function of collision centrality. In the next section, we will disentangle these two contributions -- momentum broadening and energy loss -- to the value of the variance by utilizing the momentum imbalance of $D\overline{D}$ pairs. An alternative way to distinguish these two contributions -- using the half width at the half maximum -- was discussed in Ref. \cite{Nahrgang:2013saa}.

\section{Momentum imbalance of $D\overline{D}$ pairs}
\label{sec:pAsym}

\begin{figure}[tb]
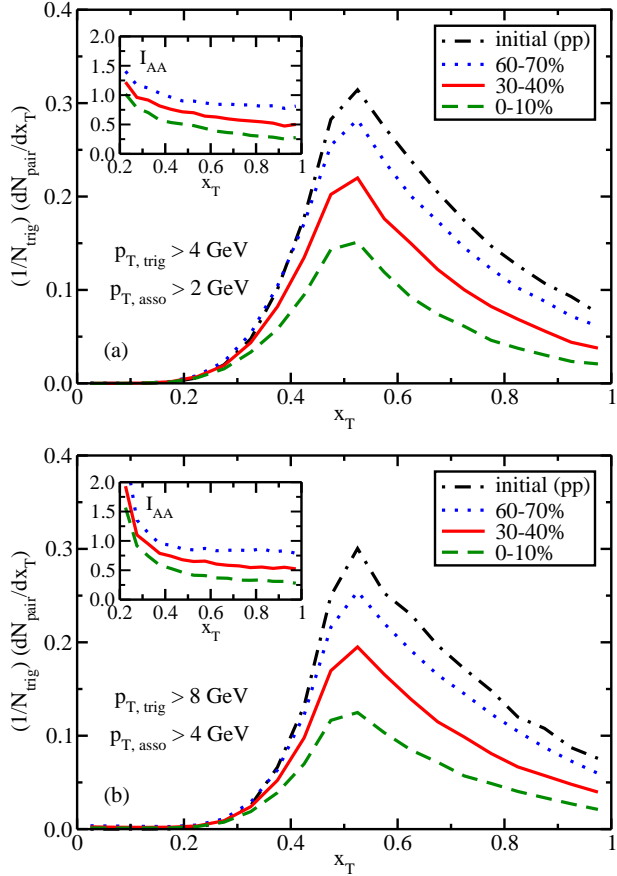

 \subfigure{\label{fig:plot-pAsym-42pt}
      \epsfig{file=plot-pAsym-42pt-wIAA.eps, width=0.45\textwidth, clip=}}
 \subfigure{\label{fig:plot-pAsym-84pt}
      \epsfig{file=plot-pAsym-84pt-wIAA.eps, width=0.45\textwidth, clip=}}
 \caption{(Color online) $D\overline{D}$ transverse momentum imbalance distribution for different centralities and different $p_\mathrm{T}$ cuts.}
 \label{fig:pAsym-DD}
\end{figure}

\begin{figure}[tb]
  \epsfig{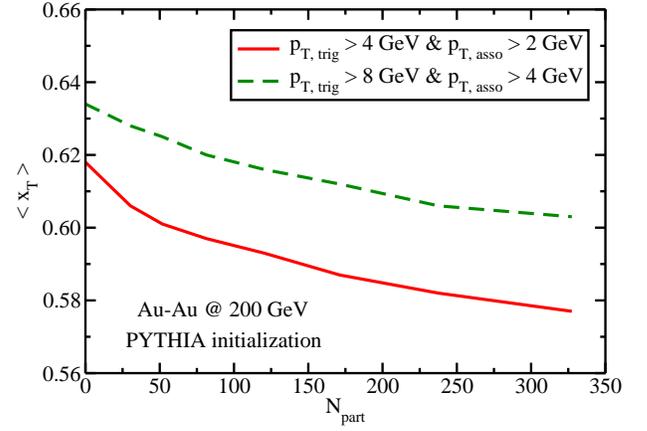}
  \caption{(Color online) The average values of $D\overline{D}$ momentum imbalance distribution as a function of the participant number for different $p_\mathrm{T}$ cuts.}
 \label{fig:plot-pAsym-npart}
\end{figure}

In addition to the nuclear modification of heavy flavor pair angular correlations, another interesting observable for studying the correlations of heavy mesons is their momentum imbalance. 
For this study, we may define the transverse momentum ratio: $x_\mathrm{T}=p_\mathrm{T,asso}/p_\mathrm{T,trig}$ for $D\overline{D}$ pairs, where $p_\mathrm{T,trig}$ and $p_\mathrm{T,asso}$ are the transverse momenta of the leading and subleading mesons, whose azimuthal angles are denoted as $\phi_{\rm trig}$ and $\phi_{\rm asso}$. For each event, we select the $D$ or $\overline{D}$ meson that has the highest transverse momentum as the leading (trigger) meson. On the back side, we look for the anti-particle ($\overline{D}$ or $D$ meson) with the highest transverse momentum and select it as the subleading (associated) meson. An angular cut is also applied for the away-side subleading meson: $|\phi_{\rm asso} - \phi_{\rm trig}| \ge 2\pi/3$; this should be helpful to reduce the background contribution and select more truly correlated pairs. 

In Fig. \ref{fig:pAsym-DD}, we show the event distribution (normalized to one  $D$ or $\overline{D}$ trigger) of the $p_\mathrm{T}$ ratio ($x_\mathrm{T}$) of $D\overline{D}$ pairs for different centrality bins and kinematic cuts. The results are shown for two different $p_\mathrm{T}$ cuts: $p_\mathrm{T,trig}>4$~GeV, $p_\mathrm{T,asso}>2$~GeV in Fig. \ref{fig:pAsym-DD}(a), and $p_\mathrm{T,trig}>8$~GeV, $p_\mathrm{T,asso}>4$~GeV in Fig. \ref{fig:pAsym-DD}(b). Two observations can be found in Fig. \ref{fig:pAsym-DD}: as one moves from proton-proton collisions to peripheral and  subsequently to central Au-Au collisions, (1) a smaller number of $D\overline{D}$ pairs per triggered event can be seen, and (2) the $x_\mathrm{T}$ distributions for both sets of $p_\mathrm{T}$ selections tend to shift to the region of smaller $x_\mathrm{T}$ (i.e., larger momentum imbalance). These are both due to stronger energy loss of heavy quarks inside the dense nuclear medium in more central collisions. As discussed earlier, there should be very few uncorrelated $D\overline{D}$ pairs contributing to the correlation functions here with the chosen $p_\mathrm{T}$ cuts. In the subfigures of Fig. \ref{fig:pAsym-DD}, we also present the ratios between nucleus-nucleus collisions and proton-proton collisions, i.e., $I_\mathrm{AA}$ as defined by 
\begin{equation}
I_\mathrm{AA}\equiv\frac{\frac{1}{N_\mathrm{trig}}\frac{dN_\mathrm{pair}}{dx}\big|_{AA}}{\frac{1}{N_\mathrm{trig}}\frac{dN_\mathrm{pair}}{dx}\big|_{pp}}.
\end{equation}
One can see that the values of $I_\mathrm{AA}$ are above unity at small $x_\mathrm{T}$ and below at large $x_\mathrm{T}$. At large $x_\mathrm{T}$, it decreases with the centrality of the collisions.

To have a cleaner description of the transverse momentum imbalance of $D\overline{D}$ pairs as a function of collision centrality, we show in Fig. \ref{fig:plot-pAsym-npart} the average value of the $p_\mathrm{T}$ ratio (or $x_\mathrm{T}$) between the subleading and the leading mesons as a function of $N_{\rm part}$ for the two sets of kinematic cuts, where $N_\mathrm{part}=0$ corresponds to the proton-proton reference. 
One can see that as one moves from proton-proton collisions to central Au-Au collisions, the transverse momentum imbalance of $D\overline{D}$ pairs increases ($x_\mathrm{T}$ value changes about 6\%). It might be easier to observe a larger medium modification to the momentum imbalance of $D\overline{D}$ pairs at the LHC with even higher $p_\mathrm{T}$ cuts, as shown in Ref. \cite{Uphoff:2013rka}. 

\begin{figure}[tb]
 \epsfig{file=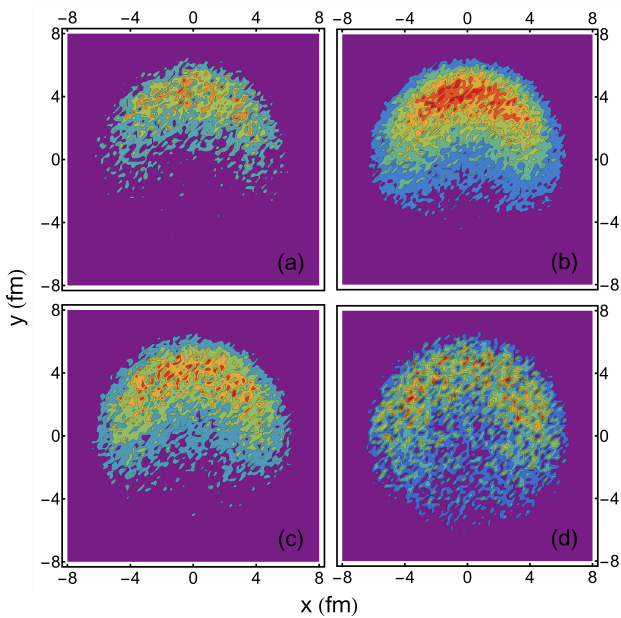, width=0.46\textwidth, clip=}
 \caption{(Color online) The density distribution of the initial $c\overline{c}$ production points ($x_\mathrm{init}$, $y_\mathrm{init}$) when the triggered $D$ or $\overline{D}$'s are taken along the out-of-plane directions ($|\phi_\mathrm{trig}-\pi/2| < \pi/6$) for 0-10\% Au-Au collisions at 200~GeV ($p_\mathrm{T, trig}>4$~GeV and $p_\mathrm{T, asso}>2$~GeV). Different values of $x_\mathrm{T}$ of the final $D\overline{D}$ pairs are compared: (a) $x_\mathrm{T}\in[0.2, 0.4]$, (b) $x_\mathrm{T}\in[0.4, 0.6]$, (c) $x_\mathrm{T}\in[0.6, 0.8]$, and (d) $x_\mathrm{T}\in[0.8, 1.0]$.}
  \label{fig:plot-initXY-xT}
\end{figure}

With the knowledge of $x_\mathrm{T}$ one may further classify the events within the same centrality bin. It has been shown in Ref. \cite{Qin:2012gp} that for $\gamma$-jet larger $x_\mathrm{T}$ corresponds to events in which hard partons are produced at the edges of the QGP fireballs while smaller $x_\mathrm{T}$ values are contributed by those pairs produced at the centers. In Fig. \ref{fig:plot-initXY-xT} we implement a similar study to investigate the correlation between the $x_\mathrm{T}$ values of $D\overline{D}$ pairs and the production positions of the initial $c\overline{c}$ pairs. We observe that for heavy flavor pairs, smaller $x_\mathrm{T}$ corresponds to events initially produced at the edge of the QGP fireballs in which one heavy quark travels outside the medium without much interaction while its partner traverse the whole QGP fireball and loses significantly amount of energy. For larger $x_\mathrm{T}$ values, initial $c\overline{c}$ pairs are more likely to spread out smoothly over the QGP. Our study thus confirms the pattern found in \cite{Gossiaux:2009mk} (Fig. 9). This allows us to utilize this momentum imbalance to probe different regions of the dense nuclear matter and gain knowledge on the path length dependence of parton energy loss as well. Note that in Fig. \ref{fig:pAsym-DD}-\ref{fig:plot-initXY-xT}, we present our calculations with both collisional and radiative energy loss switched on. We have verified that since the momentum imbalance is directly related to the total amount of energy loss of heavy quarks, its value does not strongly depend on the detailed energy loss mechanism as long as the transport coefficient is properly adjusted to describe the $D$ meson $R_\mathrm{AA}$.

\begin{figure}[tb]
  \epsfig{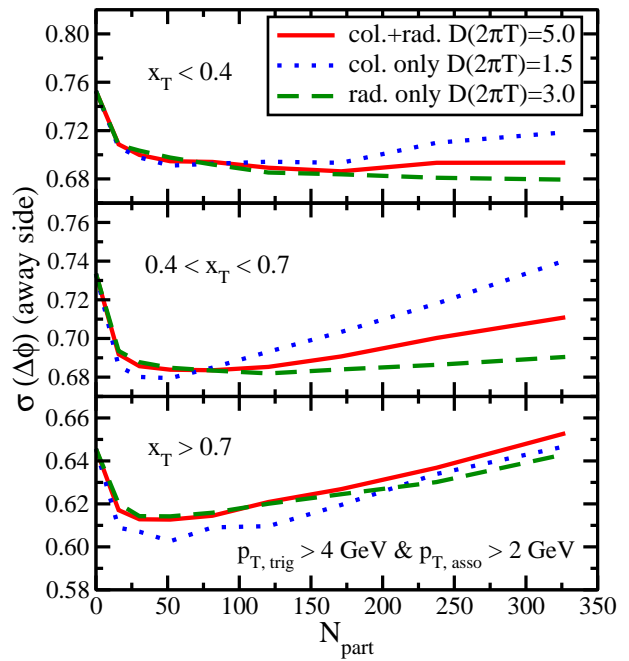}
  \caption{(Color online) The $x_\mathrm{T}$ dependence of the away-side variances of the mid-$p_\mathrm{T}$ $D\overline{D}$ angular correlation functions. Different energy loss mechanisms are shown for comparison.}
 \label{fig:plot-var2-DDbar-npart-midPT}
\end{figure}

The momentum imbalance allows us to refine the previous study on the angular correlation function in Sec. \ref{sec:correlation}. In order investigate the competition between the two ingredients that affect the variance of the angular correlation function -- momentum broadening and energy loss -- as shown in Fig. \ref{fig:var-DD}, the values of the variance are now calculated for different $x_\mathrm{T}$ bins. Here we utilize the away-side variance obtained with the intermediate $p_\mathrm{T}$ triggers as an example, which has the best potential to constrain the different energy loss models. In Fig. \ref{fig:plot-var2-DDbar-npart-midPT} we can observe when $x_\mathrm{T}$ is small ($<0.4$), it is easier for the large energy loss to overwhelm the effect of the momentum broadening. With the pure radiative energy loss, the away side variance decreases monotonically; to the contrary, the pure collisional energy loss would increase this variance in more central collisions since it smears the direction of heavy quarks more effectively inside the QGP; and the combined energy loss mechanism leads to a relatively flat variance when the participant number is large. However, in higher $x_\mathrm{T}$ bins (or with smaller energy loss), the effect of momentum broadening starts to dominate when $N_\mathrm{part}$ is large and thus increases the away side variance towards the value of a uniform distribution. And the turning point between the decrease and increase of the variance starts earlier when a higher the $x_\mathrm{T}$ cut is applied. Thus, the combination of the momentum imbalance and the angular correlation function may provide us a quantitative tool to study the detailed energy loss dynamics of heavy quarks inside the QGP.

\section{Conclusions}
\label{sec:conclusions}

In this work, we have studied the correlations of heavy meson pairs in relativistic heavy-ion collisions. 
The transport of open heavy quarks inside a dense nuclear matter is simulated utilizing our modified Langevin equation that incorporates both quasi-elastic scattering and medium-induced gluon radiation. 
The spacetime evolution of QGP fireball produced in Au-Au collisions at RHIC is simulated with a (2+1)-dimensional viscous hydrodynamic model. 
A hybrid model of fragmentation and heavy-light quark coalescence is applied to simulate the hadronization of heavy quarks into their mesonic states.

We have explored two relatively new observables for the nuclear modification of heavy flavors: the angular correlations and the transverse momentum imbalance of heavy meson pairs. 
Our results have shown that the transverse momentum imbalance of heavy meson pairs provides another measure of the energy loss of heavy quarks inside the QGP. Compared to the nuclear modification of single inclusive meson spectrum, one of the advantages of this observable is that it is not only able to quantify the total energy loss of the heavy quark, but also provides an opportunity to help probe different areas of the QGP fireball. It is found that with the increase of collision centrality, the average $p_\mathrm{T}$ ratio between the subleading and the leading heavy flavor mesons decreases, which indicates a larger imbalance caused by the larger energy loss of heavy quarks. One the other hand, the angular correlation of $D\overline{D}$ pairs provides a good candidate to characterize the momentum broadening of heavy quarks inside the QGP, and is also sensitive to different energy loss mechanisms. We find that while purely radiative energy loss does not affect very much the angular distribution between $D\overline{D}$ pairs, the collisional energy loss is very effective in randomizing the propagation direction of heavy quarks, especially at low to intermediate $p_\mathrm{T}$. To better quantify the medium modification of the heavy flavor pair correlations, we have also calculated the variances for both near-side and away-side $D\overline{D}$ angular distributions and investigate them as a function of collision centrality and transverse momentum cuts. By utilizing different bins of the momentum imbalance, the competition between the two effects -- momentum broadening and energy loss -- on the medium modification of the variance can be clearly seen on the away side when intermediate $p_\mathrm{T}$ cuts are implemented.

Our study constitutes an important contribution towards a more quantitative understanding of the energy loss and momentum broadening of heavy quarks inside a hot and dense nuclear matter.
If heavy flavor meson pair correlations are measured in the future, one can not only infer how much energy is lost from heavy quarks, but also gain some insights on the detailed interaction mechanism between heavy quarks and the QCD medium; the later seems quite difficult by only studying the nuclear modification of single inclusive heavy flavor meson spectra. Nevertheless, it is worth noticing that in order to draw further quantitative conclusion on the heavy flavor dynamics from these correlation functions, it is crucial to develop a sound method for background subtraction, especially in the low $p_\mathrm{T}$ regime and in central collisions.

\section*{Acknowledgments}

We are grateful to valuable discussions with X.-N. Wang and M. Nahrgang, and computational resources provided by the Open Science Grid (OSG). This work is funded by the Director, Office of Energy Research, Office of High Energy and Nuclear Physics, Division of Nuclear Physics, of the U.S. Department of Energy under Contract Nos. DE-AC02-05CH11231 and DE-FG02-05ER41367, and within the framework of the JET Collaboration, and by the Natural Science Foundation of China (NSFC) under Grant No. 11375072.

\bibliographystyle{h-physrev5}
\bibliography{SCrefs}

\end{document}